\documentclass{acm_proc_article-sp}

\usepackage{subfigure}

\begin{document}

\title{On-Demand WebRTC Tunneling in Restricted Networks}
%
%
%
%
%

\numberofauthors{3} 
%
\author{
Thomas Sandholm, Boris Magnusson, Bj{\"o}rn A. Johnsson\\
\\
\affaddr{Department of Computer Science, Lund University, Sweden}\\
\email{\{thomass,boris.magnusson,bjorn\_a.johnsson\}@cs.lth.se}
%
%
}

\maketitle
\begin{abstract}
In this paper we present the implementation of a WebRTC gateway service that can forward 
ad-hoc RTP data plane traffic from a browser on one local network to a browser 
on another local network.
The advantage compared to the existing IETF STUN (RFC 5389),
TURN (RFC 5766) and ICE (RFC 5245) protocols is that it does not 
require a public host and port mapping for each participating local host,
and it works with more restrictive firewall policies.  
WebRTC implements ICE which combines STUN and TURN 
probes to automatically find the best connection
between two peers who want to communicate. 
In corporate networks, simple hole punching and NAT traversal techniques 
typically do not work, e.g. because of symmetric NATs. 
Dynamic allocation of ports on an external 3rd party 
relay service is also typically blocked on restricted hosts. 
In our use case, doctors at hospitals can only access port 80 through the hospital 
firewall on external machines, and they need to communicate with patients who are typically behind
a NAT in a local WiFi network. VPN solutions only work for staff but not between patients and staff.
Our solution solves this problem by redirecting all WebRTC traffic through a gateway service on the 
local network that has a secure tunnel established with a public gateway. The public gateway redirects
traffic from multiple concurrent streams securely between local gateway services that connect to it. 
The local gateways also communicate with 
browsers on their local network to mimic a direct browser-to-browser connection without having to change the
browser runtime. We have demonstrated that this technique works well within the hospital network and arbitrary 
patient networks, without the need for any individual host configuration. 
In our evaluation we show that the latency overhead is 18-20 ms for each concurrent stream added to
the same gateway service, which is not discernible with a naked eye until you have more than 10
concurrent streams.    
\end{abstract}

\category{H.4}{Information Systems Applications}{Miscellaneous}


\section{Introduction}
Audio and video conferencing services have long been the ultimate communication tool between
remote participants connected to the Internet. Innovation has, however, been curbed by a plethora
of incompatible, proprietary protocols and tools controlled by a few companies. Custom
solutions have been very costly and complex to implement. At the same time the Web has evolved
from being a simple on-line newspaper to become a full-fledged operating system.

With WebRTC~\cite{webrtc} this operating system is now also 
capable of running standards-based
audio and video conferencing applications. The same flood of innovation that we 
have grown used to with the Social Web is now finally also available to audio
and video streaming applications. However, the uptake of the technology is restricted
by the assumption of a direct peer-to-peer connection between participating parties.
In particular in corporate networks this is a big issue, and has been an issue for 
traditional applications such as Skype too. The difference here is that at least the
control plane of the streaming applications can get through the firewall as standard port 80
Web traffic. The remaining problem is how to route the data plane, i.e. the RTP and RTCP 
(IETF RFC 4571~\cite{rfc4571} and 3711~\cite{rfc3711}) packets.
WebRTC mandates that these packets should go through a single port which helps to some extent, but
there is still an issue of traversing NATs and Firewalls.
The WebRTC solution to this problem is ICE RFC 5245~\cite{rfc5245} (Interactive Connectivity Establishment).
It is a bundle of a number of techniques including STUN~\cite{stun} (Session Traversal Utilities for NAT)
and TURN~\cite{turn} (Traversal Using Relay NAT) to find IP address and port pairs that can communicate directly.
STUN techniques are typically blocked because internal hosts are simply not reachable on routable IP
addresses. TURN on the other hand assumes that you can dynamically allocate ports freely on external networks that
can relay the traffic. This means that you would need to add a firewall exception for each stream,
which is complicated by the fact that these ports are dynamically allocated and can be shared among many users 
over time. 
For many corporate and hospital networks there is a very limited range of outbound
ports that may be allocated, in the most restrictive case only port 80 is allowed. 

Our solution to this problem is to leverage a two-hop, secure tunnel pipeline between two 
arbitrary networks that channels all RTP and RTCP traffic between these networks, while only 
requiring port allocations on each local network. To accomplish this we implement a 
gateway service that supports the 
ICE and STUN protocols to mimic a Web browser peer. No modifications were necessary to the browser
source code~\footnote{the WebRTC runtime used in the Google Chrome browser that we tested with 
is open source, and thus any component embedding this runtime should work well with our solution}. 
We only need to intercept WebRTC signalling messages during connection establishment.
These signalling messages must be intercepted in any case, since WebRTC does not define a standard channel
for relaying these messages, although a web socket service is commonly used. We need to extract
the credentials for the session from SDP~\cite{rfc2327} (Session Description Protocol) offers and answers as well as the 
ICE candidate IP address and port that the peers want to receive traffic on. 
Then we allocate sessions on-demand and open up UDP ports on local
gateways with lease-based tunnels through a central gateway. 
These local gateways do not need any configuration except for
pointing to the right central gateway, and  only one local gateway needs to run for each local network where 
you want to run WebRTC browser clients. Similarly, the browser client needs to point to the GUID of the local gateway.
Except for those configurations, the session setup and tear-down is fully automatic.

We evaluated the approach by deploying it in a real hospital network to tunnel traffic to different clients behind NATs.
The measured overhead in our experiment is less than 20ms.

This work was done as part of a project to provide IT-based support
for home-care treatment of cancer 
patients~\cite{itacih} in Southern Sweden.

Our contributions in this paper include:
\begin{itemize}
\item the design and implementation of a gateway service capable of tunneling SRTP media traffic compatible with ICE and WebRTC; and
\item a set of experiments to measure the latency, session setup overhead, and multi-stream scalability of our solution.    
\end{itemize}

The rest of this paper is organized as follows. In Section~\ref{sec:related} we present related work. 
Section~\ref{sec:home} discusses the requirements on our work from the home-care project.
In Section~\ref{sec:rtc} we give an overview of WebRTC and in Section~\ref{sec:tunnel} 
we provide some background information on a pervasive computing middleware toolkit, called Palcom, that
is leveraged in our work. Then in Section~\ref{sec:firewall} we present the design of our gateway. 
Our solution is evaluated in Section~\ref{sec:evaluation}, and then we conclude in Section~\ref{sec:conclusion}.

\section{Related Work}\label{sec:related}
In ~\cite{vline} the general problem of
sending WebRTC UDP traffic across corporate firewalls
is discussed and why current TURN and STUN solutions
do no help. Their tunneling solution does not solve the
port allocation problem, and does not allow load balancing
traffic across gateways on the local network. All traffic
also goes through their central server, whereas our solution
allows distributed central gateways as well. 

Loreto et al.~\cite{loreto2012} discuss how multiple media streams can
be sent through the same port for the same session with WebRTC today to
avoid ``opening a new hole for each stream used''. However,
this solution does not allow you to send multiple streams from different sessions
through the same port.

Rodriguez et al.~\cite{alonso2013} proposed a Multipoint Control Unit (MCU) 
architecture for WebRTC to provide additional services such as multi-party
conferences, gatewaying between protocols and call recording. They, however, do not
address the specific challenges of firewall traversal in restricted networks.

The SIP community has been heavily involved in the development of WebRTC and thus the
current standard allows for seamless tunneling between SIP gateways and WebRTC browsers
as described in~\cite{amirante2013}. Architecturally, the SIP gateway solution is similar,
but servers such as Meetecho RTCWebLite~\cite{amirante2013} focus more on protocol and signalling plane
conversions rather than the firewall traversal problem that we address.


Predating WebRTC, 
a large number of firewall traversal solutions have been evaluated
for relaying SIP-based audio and video sessions through Enterprise networks in~\cite{stukas2004,chatterjee2005,ipfreedom}.
The techniques studied include STUN, UPnP, MIDCOM, Sen, FANTOM, STEM,
PSTN Gateway, SIP Proxy, Application Level Gateway, DMZ MCU, 
and Semi-Tunnels/Transparent Traversal. 
The last technique in the list~\footnote{implemented in a product called IPFreedom by Ridgeway Systems}~\cite{ipfreedom}, is similar architecturally to our approach in that a 
local network client relays multiple
sessions through a public network server. The motivation in that case was to avoid having to change legacy 
firewall deployments, while we also have this goal our main motivation for this architecture is to
avoid having to change the browser source code.

Our solution differs from all of these pre-WebRTC solutions as they do not leverage a signalling plane (in our case WebRTC JSEP) that trivially traverses firewalls via port 80/443. Furthermore our solution only requires a standard browser to be run on a participating client machine.

\section{Home care requirements}\label{sec:home}
The work in this paper is motivated by the requirements from a 
project in applied research involving the University hospital in Lund, Lund 
University and 6 companies~\cite{itacih}. The aim is to develop IT-support for 
the situation where patients are enrolled at the hospital, but given advanced 
care at home. This is a form of care that is growing rapidly, in particular for 
chronically ill patients, and seen as a benefit both for patients and the 
hospital, as long as the patients feel safe in this situation. It does, however, 
demand new kinds of support systems. Medical equipment at home needs
 to be remotely monitored (and controlled) from the hospital in order to 
avoid frequent technically-motivated visits to the home. The mobile staff 
needs new facilities for handling the information flow in order to avoid 
doing the same task twice (notes on paper later entered into a computer system), having 
updated information at hand, and work efficiently in general. Communication 
over a variety of media, including video, between patient and staff is 
believed to be crucial for many patients to feel safe in this situation. 
Existing video communication systems, such as Skype and Lync, are not  
applicable in this situation due to security concerns,such as
\begin{itemize}
\item access to the patient on video needs to be restricted, 
\item management overhead,
\item separate registration of patients addresses, 
\item and lack of integration possibilities.
\end{itemize}
We aim at a 
single integrated support system for home care. With these concerns, the 
possibilities offered by WebRTC (see Section~\ref{sec:rtc}) are very attractive. 

\begin{figure}[htp]
\centering
\includegraphics[scale=0.3]{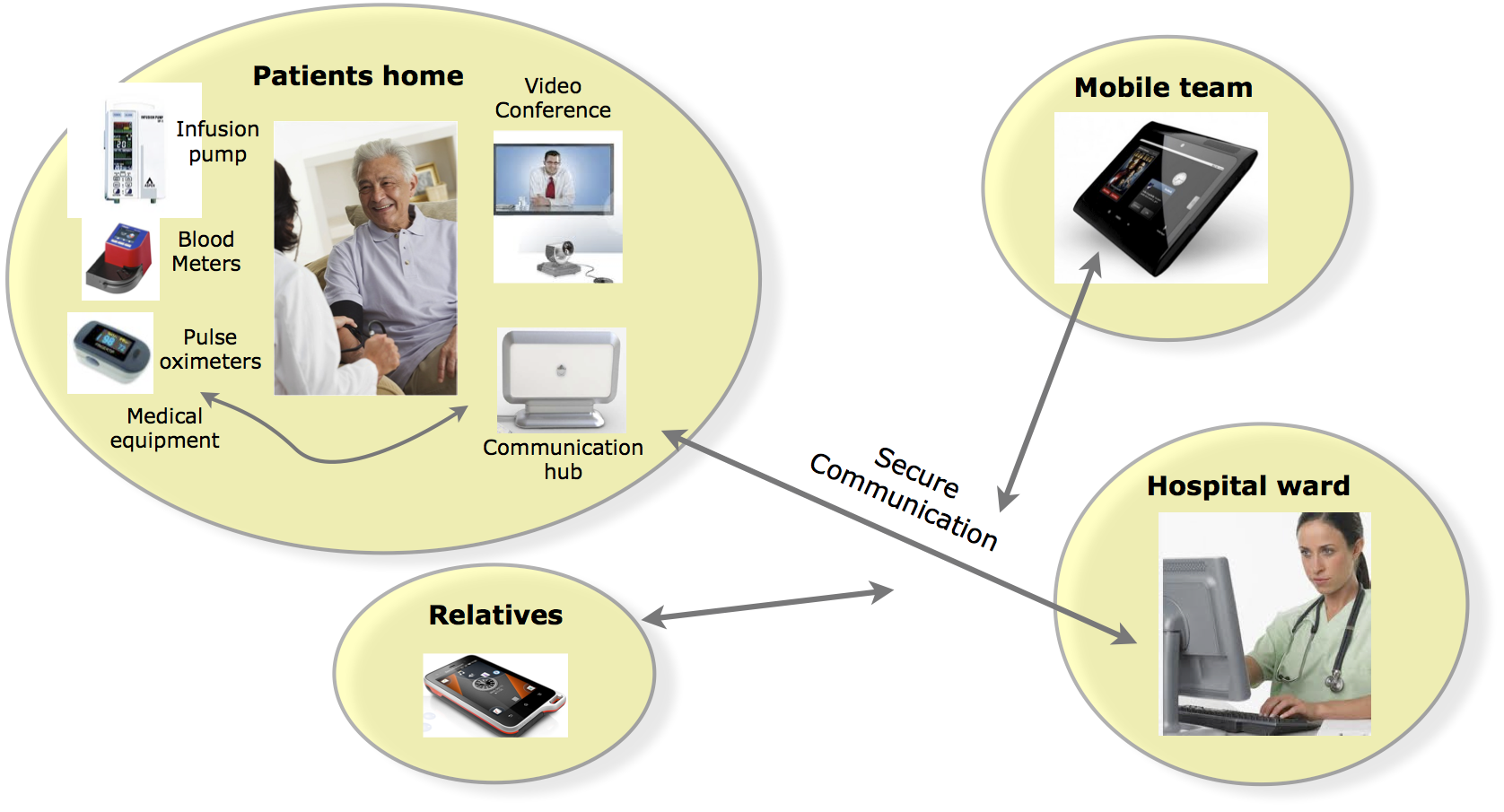}
\caption{The itACiH infrastructure for home care.}\label{fig:itacihOverview}
\end{figure}

The technique demanded from a security point of view  in a healthcare system 
for remote access by the medical personnel  (VPN-tunnels with 2-factor 
authentication using hard certificates on  identity cards)  is not practical
to use for connecting patients and home-based equipment. The architecture 
chosen is instead based on having  servers outside 
the hospital firewall, and connect out from inside the hospital, and thus being in full 
control of if and when to connect and what information to make  
available outside the firewall. This is an architecture that is also preferred from 
a security point of view by the hospital over potentially 1000s of constantly 
open VPN tunnels. The very restricted policy of which ports and 
protocols to open in the firewall, lead to the conclusion that we need to 
communicate using TCP/IP over port 80. With the communication software we 
are using, all logical connections (supporting the need as outlined above) are 
multiplexed over one such link. When adding support for video in this set-up, we thus 
also want to multiplex the video over this link. Network-wise, we assume that 
this link is realized on a high-speed network - perhaps the computers involved 
are even physically in the same computer room. The bandwidth should thus not be 
a problem here even if there are several video-connections active at the same time.

The equipment in the other end, in the patient's home, or in the field carried by mobile nurses, 
are using a similar multiplexing setup, but must rely on much weaker infrastructure, such 
as 3/4G mobile technology or slow up-links from the patient's home.

\section{WebRTC}\label{sec:rtc}
WebRTC is a new standard being developed by W3C~\cite{webrtc} (browser API) and IETF~\footnote{http://tools.ietf.org/wg/rtcweb/} (wire protocol) 
to allow audio and video
conferencing capabilities without plugins in Web browsers. This means that
audio and video-streaming applications can be developed directly in Javascript
and deployed and run through a standard Web browser. This approach is
is very attractive for our scenario described in Section~\ref{sec:home}, because most devices
today run a Web browser. Furthermore, in certain restricted environments, such as a
hospital, a Web browser is the  {\it only} end-user application that is allowed to run. 

The WebRTC API comprises two main parts. First, the HTML5 GetUserMedia (GUM) API~\cite{gumapi},
is used to give the Web page access to media from local audio and video capturing devices. For video,
the picture frames may, for instance, be mapped to an HTML5 Canvas or Video element for direct pixel-by-pixel access
and display respectively. 
Second, the PeerConnection API allows these video and audio streams to be attached to a peer-to-peer
audio and video-conferencing session.
The connection is established through an Offer and Answer protocol, based on SDP negotiation, 
that is transmitted over a signalling channel. This protocol is defined in an IETF draft
called Javascript Session Establishment Protocol (JSEP)~\cite{jsep}. It allows the
Javascript application to intercept and enhance the session parameters being negotiated.
This protocol is what allows our solution to redirect the traffic through our gateway without
having to modify the browser. The fact that these APIs are being standardized also means that
the code we use to intercept and modify the session establishment is generally applicable
across browsers in theory~\footnote{as of this writing only Chrome and Firefox support the latest WebRTC 
PeerConnection and JSEP APIs, and minor differences in the Firefox implementation has forced us
to restrict testing of our solution to Chrome.}. One key part in the use of JSEP for our scenario
is that the signalling plane communication is not specified, which means that the browser just gives
the application callbacks with payloads, e.g. SDP, to send to the other peer. Conversely,
when a WebRTC payload is received at the other end it is fed into the WebRTC runtime by calling
APIs on the PeerConnection API after potentially having read and modified it. 
The only requirement on this application-level channel is that it needs to be real-time, and there
needs to be a way for peers to rendezvous.
Hence, many applications use COMET~\cite{COMET} or WebSockets~\cite{rfc6455} for this purpose.
 
However, for the data plane, the suite of WebRTC APIs assume that the application does not
intercept or modify the communication. The UDP packets are sent on internal sockets
directly between the browser WebRTC runtimes.
This is the main issue we need to solve as the current PeerConnection API 
assumes that there is a direct connection between the peers (that can be established through ICE~\cite{rfc5245}) .
As previously discussed, in most corporate networks including hospital networks, it is not possible to establish
such direct connections to peers outside the local network.

The way to solve this particular issue of also redirecting the data traffic is to 
intercept and rewrite the ICE endpoint candidates sent using the JSEP protocol. 
We will discuss in more detail how we accomplish this in Section~\ref{sec:firewall}.

An alternative solution could have been to plug in our tunnel behind a custom server supporting
the TURN protocol and use the standard JSEP exchange without modifications. However,
we opted not to pursue this approach at this point since it would involve implementing the full TURN
stack and protocol, when we already have most of this functionality in our existing middleware, Palcom, which we describe next.

\section{Palcom}\label{sec:tunnel}
The Palcom architecture~\cite{palcom} includes Palcom devices (or applications) which 
can communicate on a local network and tunnels that connect Palcom 
devices over remote networks and multiplex their communication 
over a single TCP/IP port. Since every Palcom device can act as a 
router of Palcom discovery and application traffic, one such Tunnel is 
enough for all Palcom devices on one side of the Tunnel to connect to 
any of the Palcom devices on the other side of the Tunnel (and vice versa). 
If Palcom Tunnels are established from Palcom devices on local networks 
behind NAT or Firewalls to the same Palcom device on a visible network 
it appears to the Palcom devices as if they are on the same local 
network when it comes to discovery and communication. 

The Palcom tools include an application that serves as a service container,
and is exposed as an empty device, called ``theThing''. 
This application can dynamically load and execute functionality in terms 
of Palcom Services. The 
functionality we used in our WebRTC gateway is implemented as such 
Palcom Services and deployed to instances of theThing executing in 
local networks and a central location. The Tunnels are provided by 
the instances of theThing and can swiftly be configured to use whatever 
ports that might be open. This architecture thus clearly separates the 
functionality that talks the WebRTC protocols (embedded in the 
Palcom Services) from the functionality that provides the NAT and 
Firewall traversal techniques. Tunnels used in this work are built 
on TCP/IP in order to get through tight firewalls, where one can expect that 
the only option is to use port 80. UDP is in general much more efficient
and would be attractive to use in less restricted situations. In this work all 
measurements has, however, 
been done using TCP/IP based tunnels. An interesting alternative would 
be to use TCP for the Hospital-Server link (typical over a high-speed 
wired network) and using an UDP based tunnel for the Patient-Server 
link (typical over wireless slow up-link networks). The  Palcom architecture
would handle such a situation without change since the general
communication and routing mechanisms are independent of the network
technology used. This is made possible because Palcom includes a message 
transport protocol that can encapsulate data of any application protocol 
(such as RTP and RTCP in this work).

Palcom services are also readily accessible from a Web browser thanks to
a recently developed Web bridge~\cite{palcom2012webreport}, which makes them
a good fit for our WebRTC gateway implementation.

Next, we describe how we implement our WebRTC gateway using Palcom, and
the WebRTC APIs and protocols.

\section{WebRTC Gateway}\label{sec:firewall}
Our solution comprises three services, a tunnel (TUN), a local RTC (Real-Time Communication)
gateway (LGAT) and a central RTC gateway (CGAT). All control plane traffic 
between the components go through a Web Server using a publish and subscribe
model implemented with long-polling and exposed with a REST interface that can be
accessed through an API similar to HTML5 WebSockets~\footnote{as the standards
and implementations of WebSockets mature we can replace our custom pubsub
protocol with WebSockets}. 

The LGAT services are typically deployed within local networks behind firewalls 
and the CGAT service is deployed on the public Internet. 
The TUN service, described in more detail in the previous section, is responsible 
for multiplexing the traffic that comes into it through a single well-known port 
to another tunnel peer which would sit on a different network such as the public Internet.
One could imagine tunneling directly between the peer networks but that makes the configurations
more volatile and it may also be a security issue to give access directly into
services running in a remote local network (such as a hospital or corporate network).

To simplify deployment we by default deploy a default CGAT service 
so that only local gateway configurations need to be set to establish a session. 
If the CGAT service 
detects that both peers use the same LGAT no gateway redirection
will be performed and the standard WebRTC protocol will be used.
However, if the peers have two different LGATs we set up and communicate through
TUN in four initiation phases: offer and answer exchange, candidate exchange, STUN
endpoint verification and finally SRTP (Secure Real-Time Protocol) and SRTCP (Secure Real-Time
Control Protocol) data traffic exchange. All of these phases are present in the standard WebRTC
browser-to-browser communication as well. Next we describe in more detail how we extend each phase
to create an efficient and secure WebRTC tunnel.  

\subsection{Phase I: Offer and Answer Exchange}
\begin{figure}[htp]
\centering
\includegraphics[scale=0.35]{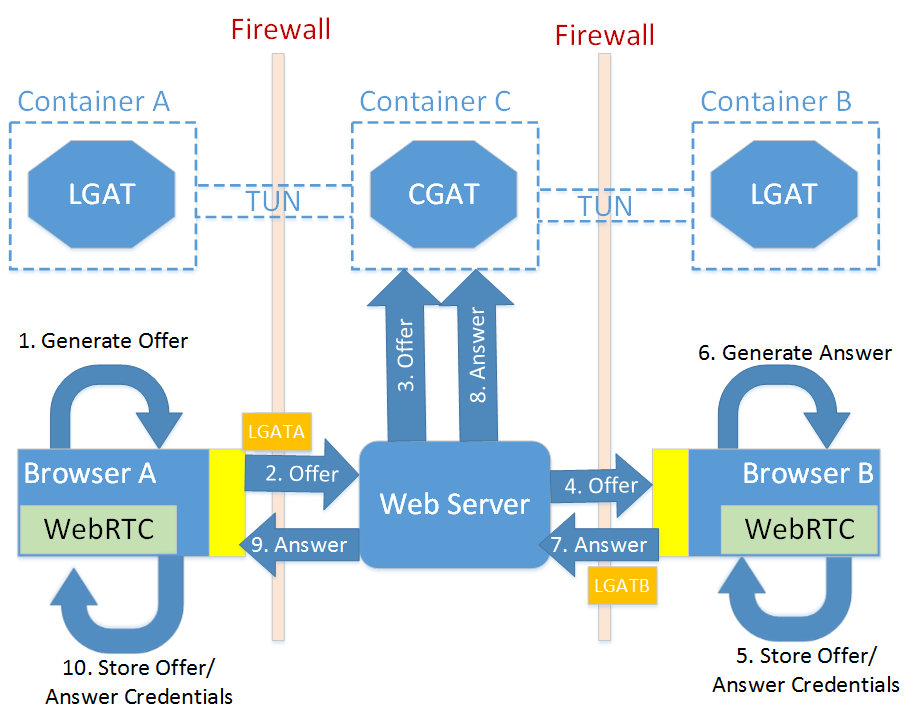}
\caption{Phase I: Offer and Answer Exchange.}\label{fig:GatewaySetup}
\end{figure}

WebRTC does not specify how peers rendezvous, or how control messages between them are relayed.
As previously mentioned, we take the some approach as most existing applications of sending the
messages through a WebSocket-like pubsub service. In this first phase we do not make any
modifications to the standard exchange protocol except to piggy back gateway data in the messages.
This phase comprises the following steps (see Figure~\ref{fig:GatewaySetup}):

\begin{enumerate}
\item The WebRTC runtime in browser A generates an SDP (Session Description Protocol) offer
      through JSEP 
      containing among other things user and password credentials for the session.
\item We intercept the offer in a local Javascript library (yellow box) before it is broadcast to potential peers.
      The signalling plane of WebRTC is not defined in the specification but could treat the JSEP protocol
      as a black box, i.e. no knowledge of payload is necessary, you just need to pass the data from
      the WebRTC runtime in browser A to browser B by some means. This makes it easy for us to attach
      additional information to the signalling payload without the knowledge of the respective
      WebRTC runtimes while still being fully compliant to the JSEP handshake protocol.
      The information we add is simply the identity of the LGAT of browser A (LGATA).
\item The enhanced offer is now passed through the CGAT service (which keeps track of who is online
      in which virtual Web conference room) and broadcast to everyone listening on
      the pubsub channel where browser A sent the offer.
\item Browser B receives the offer payload and we intercept the SDP data before it reaches the
      WebRTC runtime. 
\item We extract the credentials (ice-ufrag and ice-pwd) from the offer SDP. These credentials are
      cached locally in the Javascript runtime for later use.
\item Browser B will now generate an answer SDP to browser A's offer. We intercepts this answer too
      before sending it off to browser B, and extract the answer credentials. 
      Now by combining these credentials with the credentials extracted from the offer we have all
      the information needed for our gateway to verify incoming connection requests and to sign 
      outgoing connection requests properly. In particular the following two pairs of user name
      and passwords are constructed: $\{u_1:u_2,p_1\}$ and $\{u_2:u_1,p_2\}$, where $u_x$ is the
      ice-ufrag attribute of user $x$ and $p_x$ is the ice-pwd attribute of user $x$.
      As the answer reaches browser B the credentials will be extracted similarly there.
\item The Answer from B is sent back through the Web server.
\item It follows the reverse route through the CGAT back to browser A.
\item The Answer is received by browser A and intercepted again by our Javascript library.
\item The credentials are extracted from the Answer SDP and cached locally in browser A.
\end{enumerate}
This completes the first phase of our tunnel setup.
 
\subsection{Phase II: Candidate Exchange}
\begin{figure}[htp]
\centering
\includegraphics[scale=0.35]{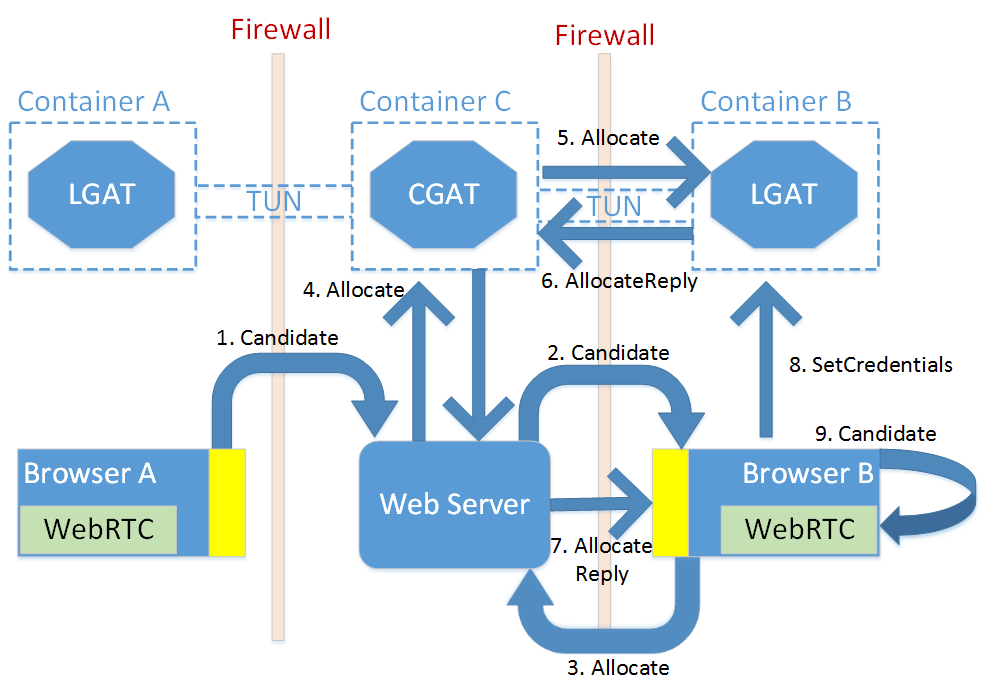}
\caption{Phase II: Candidate Exchange.}\label{fig:GatewayAllocate}
\end{figure}

The following phase of candidate IP address and protocol (TCP or UDP) exchange
may start before the offer and answer exchange has completed fully, so we need to account
for certain events in this phase sometimes happening before and sometimes after
events in the previous phase. Our general solution to that problem is to cache away results
needed for an action and when we detect that we have complete information from both
peers we fire our event to proceed to the next step.
Most of the heavy lifting of our gateway setup is done in this phase which
proceeds in the following steps (see Figure~\ref{fig:GatewayAllocate}):
\begin{enumerate}
\item The WebRTC runtime in browser A generates a candidate IP address, port and protocol. 
If browser A is on a local network it could send us a non-routable IP, such as 192.168.0.10.
This is no problem for our gateway though since we assume that LGAT is on the same local network
as the browser, which is sufficient to route traffic. We pass the (ice-)candidate without any
modifications to the Web server through the session that was established in the previous phase.
\item Browser B receives the candidate sent by browser A. Now we intercept the message before
passing it on to the local WebRTC runtime. 
\item In the previous phase, browser B received the LGAT of browser A.  This LGAT reference together with the IP, port
and protocol information that was just received are now sent to the Web server in an allocate request targeted
at the local LGAT service in container B. Note that the browser can only communicate with HTTP to the same domain
as the Web page (unless we use a JSONP hack) and can thus not send this message directly to the local LGAT.
Exposing this local container through a JSONP cross domain HTTP API could be a security vulnerability though,
which is why we can only communicate with the LGAT through the CGAT. From a developer perspective it
is also much less complicated since we can just send the messages through the same pubsub API as all
other WebRTC control messages.
\item The Allocate request goes through the CGAT, as the web server may not be o the same network as the LGAT
services, which is the whole point of having our tunnel.
\item The Allocate request is received by the local gateway. An internal session is now established and we
store the remote LGAT in Container A, as well as the port,IP and protocol (for now we only support UDP, but it
would be trivial to handle TCP as well) that was in the candidate message generated by browser A. We also
allocate a new port on the local network and send the port and IP as well as the session id back.
\item The allocate reply passes through the CGAT which takes note of the session id in order to 
block unknown sessions to be forwarded to the LGAT later on. 
\item The allocate reply then reaches Browser B again, and the allocated IP and port are extracted.
\item At this point when we know that a session has been allocated in our local LGAT service,
we can also store the credentials we cached in the previous phase in the LGAT service. Note that
this request goes through the CGAT as well for the same reasons as mentioned above. Also note that
the reason why this is a separate request as opposed to being data just piggy-backed on the allocate
request is that the first and the second phase overlap. This means that the credentials may not be available
at the time the allocate request is called. Of course we could still piggy-back it if it is available,
but we, in either case, need to have a separate operation to set the credentials out-of-band. 
\item Now the original candidate request received in step 2 is modified to contain the just allocated
IP and port instead of the IP and port allocated by browser A. Once that is done the modified candidate
is sent to the WebRTC runtime as if it was received directly from browser A. 
\end{enumerate}

This is the first step
in redirecting the WebRTC traffic to our tunnel. However, it only results in the browser sending
out STUN verification requests to the IP and port we redirected it too. In order to receive valid
data traffic on that host and port we need also respond to the STUN requests correctly, which is
the task of the next phase.  

\begin{figure}[htp]
\centering
\includegraphics[scale=0.35]{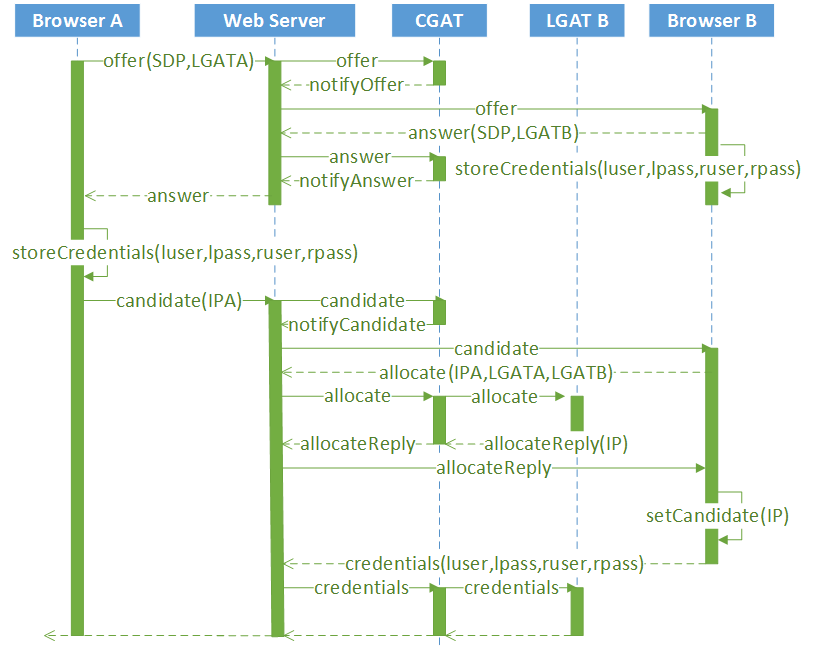}
\caption{Offer/Answer and Candidate Exchange UML Sequence Diagram.}\label{fig:GatewaySetupSequence}
\end{figure}

The parameter passing and operation sequence of the first two phases are summarized in
a UML sequence diagram in Figure~\ref{fig:GatewaySetupSequence}.

\subsection{Phase III: STUN Endpoint Verification}
The final step before being able to send data traffic on our tunnel is to verify our allocated
IP and port with the Browser WebRTC runtimes. This phase can overlap with the previous two phases
because the two peers run the phases concurrently to each other. However, we are guaranteed
that it does not happen until we set the allocated IP and  port in the candidate rewrite 
step (step 9) in the previous phase for the local browser. This means that we may not be
ready to fully answer the initial STUN requests as the browser start generating them.
There is a very sensitive balance between allocating the candidate too late and too early.
The browser gives up after sending too many unanswered STUN requests. However, it may also give
up if we do not provide any candidates soon enough. Another issue is that the browser
bombards the LGAT with STUN requests, so trying to answer them all before the setup is complete
can slow down the gateway to the point that the final setup calls can not be served.
For this reason we can configure the LGAT service to ignore a certain number of STUN
requests before attempting to answer them correctly. There is a default value that
works in most cases but for very CPU deprived machines it may need to be modified. 

\begin{figure}[htp]
\centering
\includegraphics[scale=0.35]{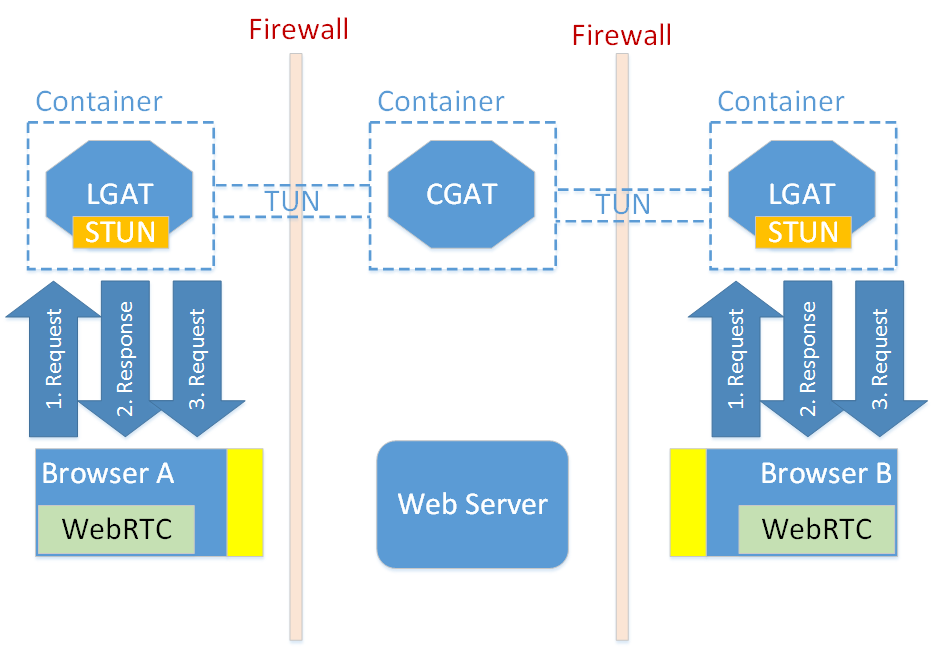}
\caption{Phase III: STUN Endpoint Verification.}\label{fig:GatewayStun}
\end{figure}

The verification proceeds concurrently in both browsers in the following steps (see Figure~\ref{fig:GatewayStun}):
\begin{enumerate}
\item The browser sends a STUN BindingRequest to the allocated IP and port using UDP. At this point
      we need to determine if it is a STUN request very efficiently since the actual data  will pass
      through the same IP and port using the same transport protocol (UDP). 
\item Once we have determined that
      it is a valid STUN BindingRequest and we have the correct user credentials to answer the request
     (and the number of STUN packets configured to drop have been dropped) we create a 
     STUN BindingResponse. The response contains the integrity attribute, which means that
     it is signed with the user credentials of the remote browser. Note that the STUN requests
     are never passed through the tunnel, we have all the information needed to answer them in the
    local LGAT, which is an optimization, as it saves network traffic.
\item However, this also means that the local browser will never receive any STUN Binding Requests, which
     is also a requirement for the data traffic to commence. Hence, we also compose a STUN BindingRequest
     at the same time as the BindingResponse is composed. Both the BindingResponse and the new BindingRequest
     are then sent on the same browser IP and port that the original BindingRequest came from.
\end{enumerate}

In earlier versions of Chrome this was a Google specific ICE protocol without message
integrity checks and message signatures. Hence, earlier versions of our services implemented
this proprietary protocol. Now since Chrome 24 both Chrome and our LGAT services will send STUN
messages according to the standard ICE RFC 5245 protocol including message integrity signatures.
We are now ready to route media streams, which is described in the next phase.

\subsection{Phase IV: Data Traffic Routing}
\begin{figure}[htp]
\centering
\includegraphics[scale=0.35]{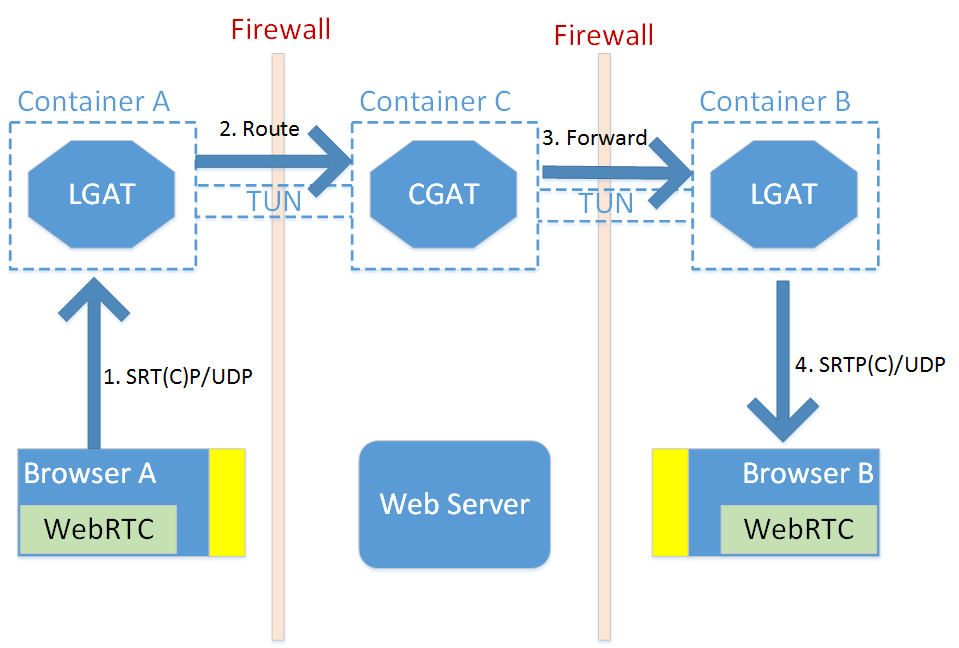}
\caption{Phase IV: Data Traffic Routing.}\label{fig:GatewayRoute}
\end{figure}

The routing of WebRTC media streams proceeds as follows (see Figure~\ref{fig:GatewayRoute}):
\begin{enumerate}
\item Browser A sends the SRTP or SRTCP packets over UDP to the IP and port we
allocated for this session in the previous phases.
\item The LGAT service then looks at the browser B LGAT service associated with this session as well as the IP and port
of the browser B. This info is sent to the CGAT service with the data packet. The reason that it is
not routed directly to the remote LGAT is that they sit on different networks and we may not want to give full
connectivity of services across sessions to, for instance, protect patient data in a hospital doctor and patient 
exchange which was the use case that motivate us to build this tunnel.
The WebRTC RTP payloads are all encrypted so one only needs to make sure that the SDP singalling plane is encrypted
where the credentials are included to secure the protocol. This means that if someone else receives a data 
packet, e.g. by sniffing our tunnel or RTC gateway traffic they cannot do anything with it. We also have
the option of encrypting our tunnel traffic, but in this case it is overkill.
To encrypt the signalling traffic we simply rely on HTTPS which is supported by our Web socket
implementation trivially since it only relies on the browser XHR (XMLHttpRequest aka AJAX) API~\cite{palcom2012webreport}.
\item The central container does a quick session verification to make sure that the browser B LGAT in fact
expects the packet before forwarding it.
\item When the data packet arrives at the remote LGAT it can be directly forwarded without any further checks.
      Optionally we can do a checksum since all that information is available, but since the WebRTC traffic is so
      sensitive to latency we avoid intercepting and processing the actual data packets more than necessary. The WebRTC
      browser runtimes drive the full SRTP and SRTCP protocol, so there is no need to reinvent that security either.
\item Finally, the data packets arrive at browser B, and will show up in the browser UI.
\end{enumerate}

\subsection{STUN Server}
As a final note we now describe our ICE STUN server inside the LGAT service in some more detail.
As previously mentioned this service is responsible for both responding properly to 
STUN messages received by the browser runtime on the local network as well as to generate
STUN messages to simulate the remote browser sending STUN messages to the local network browser.
Our ICE STUN implementation uses the ICE4J library to compose and parse STUN messages.
We also add an efficient way of checking whether the message is a STUN binding request or a RTP/RTCP
packet by peaking into the assumed STUN header, so we do not try to parse every incoming packet, as
all the packets will be sent on the same UDP IP address and port.
Now if we determine that the incoming packet is a STUN requests we parse the Transaction ID and extract the
IP and port of the sender of the packet (the browser WebRTC runtime). The IP address and port
are now used to create a XOR mapped address attribute in a binding success response.
We also add the integrity and finger print attributes according to RFC 5245. In order to properly
create these attributes we need to find a key to sign the message with. This key is obtained
from the password mapped to the ufrag derived username attribute matching the STUN username attribute of the
incoming stun binding request. 

Now half of the work is done. In addition to replying to the SUN binding request
we also need to generate our own binding request to the local browser to mimic the remote browser sending this
request. 
This is done as follows. The username of the incoming request is flipped so the part before ":" is appears
after and vice versa. Now the password of this flipped user name is looked up to sign the integrity
part of the request. Priority, Controlling and TieBreaker attributes are populated either with default
values or if they appear in the incoming request they are simply copied over. We also add an empty UseCandidate 
attribute because Chrome tends to do that. Next, both the request and response are sent to the appropriate UDP
host and port of the local browser and we are done.
We note that the integrity checks are done with short term credentials according to RFC5389 and RFC5245.
It is trivial to support long-term credentials in our gateway too as it would just involve setting
up the username and password pairs from another source than the SDP answers and offers. So only
the Javascript application code would need to be modified slightly.

\section{Evaluation}\label{sec:evaluation}
To evaluate our gateway we run experiments to measure the latency impact 
for ongoing sessions,the overhead incurred when initiating a session, and 
the scalability in terms of concurrent streams.

\subsection{Latency Evaluation}
We leverage the WebRTC getStats statistics 
API on the PeerConnection directly from Javascript in one of the
participating browsers~\footnote{Google Chrome Version 30} 
to extract stream performance values. 
Sampling was done on a second-by-second basis,
and averages were calculated each minute. We base our final metrics
on statistics from 7-minute long test sessions, where the first two
minutes are ignored as a warmup phase.  Although we extracted many
more metrics, two key metrics that we base our comparisons on are
round-trip-time (RTT)~\footnote{googRtt in stats type ssrc} and 
Jitter~\footnote{googJitterReceived in stats type ssrc}. We also
looked at loss and frame rates, but could not see any interesting 
differences between direct connections and tunneled connections
in those metrics.

Since our gateway can be deployed both as a personal gateway on
the same host as the browser and as a shared gateway in an internal
network, we evaluate the impact of local (personal) and remote 
gateways. In the remote gateway experiment we let the remote
gateway run on the same host as the remote peer. Hence we only add 
network traffic in this setup while keeping the processing power
the same. 

To further study the impact under different network settings we run
experiments using both wireless (WiFi) and a wired (Ethernet) access to the same network.

During our experimentation we also found that the performance of the peer machines
make a big difference in terms of the latency results. We
therefore also compare our results with experiments done where one laptop
was replaced with a slower laptop. Both The faster and the slower
laptop run Windows 7 and have a 1.6-1.7GHz CPU. The Windows Experience Index
for the slower laptop is 3.2~\footnote{3.2 calculations per second, 5.6
memory operations per second, and a graphics score of 4.3} and for the
faster laptop 5.6~\footnote{5.6 calculations per second, 5.9
memory operations per second, and a graphics score of 5.6}. 

\begin{figure}
\hfill
\subfigure[RTT]{\includegraphics[width=4.1cm]{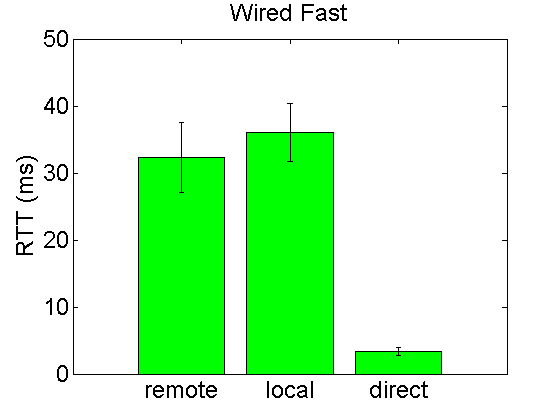}}
\hfill
\subfigure[Jitter]{\includegraphics[width=4.1cm]{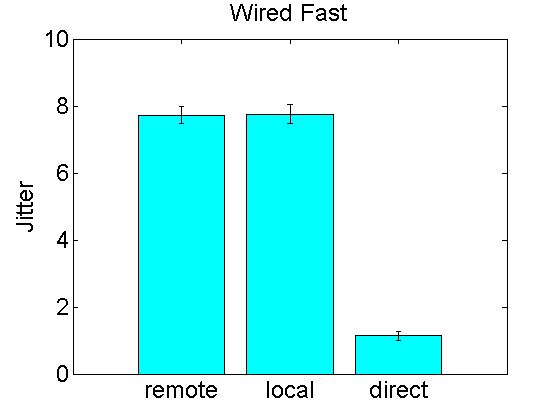}}
\hfill
\caption{Local and remote gateway performance compared to direct streaming with the fast laptop
on the wired network.}\label{fig:wiredfast}
\end{figure}

The RTT and jitter results for the fast laptop running on the wired network
with local and remote gateways is shown in Figure~\ref{fig:wiredfast}. The
direct value denotes the performance of a stream that follows the
default browser behavior and communicates directly between browsers.

\begin{figure}
\hfill
\subfigure[RTT]{\includegraphics[width=4.1cm]{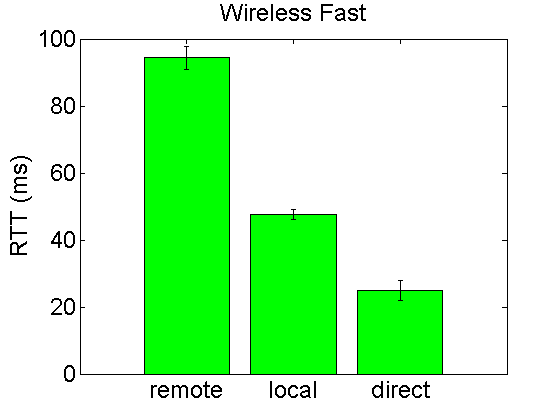}}
\hfill
\subfigure[Jitter]{\includegraphics[width=4.1cm]{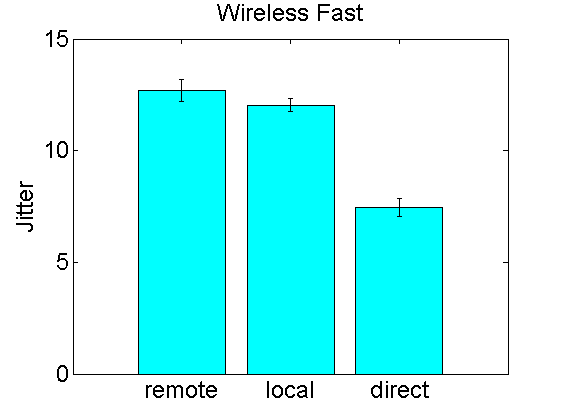}}
\hfill
\caption{Local and remote gateway performance compared to direct streaming with the fast laptop
on the wireless network.}\label{fig:wirelessfast}
\end{figure}

The RTT and jitter results for the fast laptop running on the wireless network
with local and remote gateways is shown in Figure~\ref{fig:wirelessfast}.

\begin{figure}
\hfill
\subfigure[RTT]{\includegraphics[width=4.1cm]{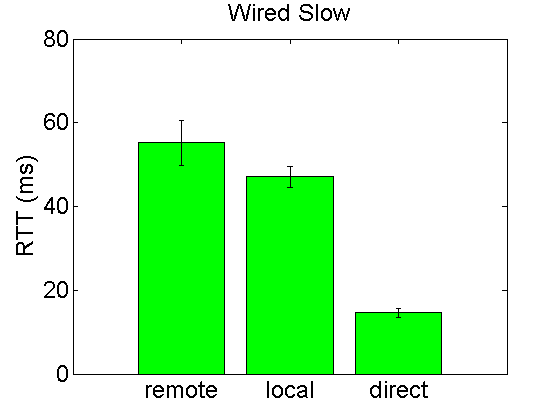}}
\hfill
\subfigure[Jitter]{\includegraphics[width=4.1cm]{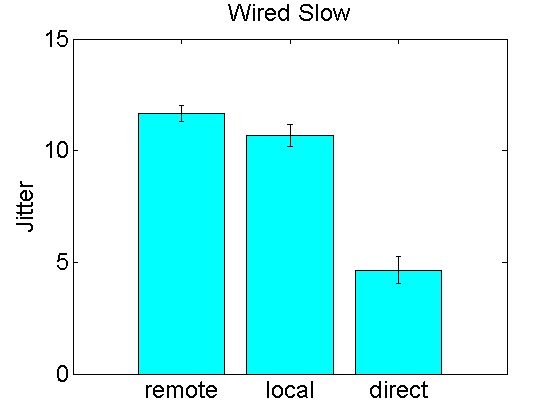}}
\hfill
\caption{Local and remote gateway performance compared to direct streaming with the slow laptop
on the wired network.}\label{fig:wiredslow}
\end{figure}

The RTT and jitter results for the slow laptop~\footnote{Note that we
only replaced one of the peers compared to the fast laptop experiments
and kept one peer (a desktop) the same.} running on the wired network
with local and remote gateways is shown in Figure~\ref{fig:wiredslow}.

Table~\ref{T:experiment} summarizes the absolute values for the RTT and jitter metrics
across the experiments.

\begin{table*}
\centering
\caption{Summary of experiment values.}\label{T:experiment}
\begin{tabular}{|l|l|c|c|c|c|c|c|} \hline
{\bf laptop} & {\bf network } & \multicolumn{3}{c|}{\bf rtt}  & \multicolumn{3}{c|}{\bf jitter} \\ 
 &  &  local gateway & remote gateway & direct & local gateway & remote gateway & direct \\ \hline
slow & wired & 47 & 55 & 15 & 11 & 55 & 5 \\
fast & wired & 36 & 32 & 3 & 8 & 32 & 1 \\
fast & wireless & 48 & 94 & 25 & 12 & 94 & 7 \\
\hline
\end{tabular}
\end{table*}

As expected, the round trip times
suffer quite a bit, but given that the alternative in many cases is no stream at all,
the slight lag may be acceptable. The frame rates are very stable and there is 
virtually no loss and only a modest increase in jitter.

For the slow laptop the local gateway does not perform so well since the machine
gets overloaded. Note that the remote gateway runs on the other peer's machine
so the overall experiment load is the same. For the fast laptop running a local gateway
is better, though, since it avoids extra cross host traffic.

\subsection{Session Initiation Overhead}
\begin{figure}[htp]
\centering
\includegraphics[scale=0.45]{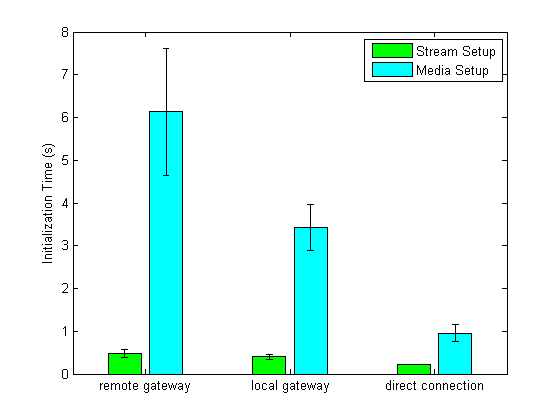}
\caption{Connection time measured as Phase I completion time (Stream Setup) 
and Phase IV start time (Media Setup).}\label{fig:connectiontime}
\end{figure}

We now study the impact on the video and audio session
initiation time. 
The overhead incurred is due to the extra redirections and remote allocations executed as described for
the four phases (I, II, III, IV)  in Section~\ref{sec:firewall}. 
We measure both the time from the user clicking on a button to initiate the call to the 
time when the Offer/Answer Exchange is done (roughly the completion of Phase I), and to
the time when the first media packet was both sent and received.  The latter part is where
the bulk of the overhead is and from a user experience perspective it shows up as the
video element on the web page being black before it starts showing a stream.
The first initiation time is easy to measure from callbacks but the second one is a
bit trickier since we do not have full access to the UDP packets inside the browser. Thus, we again rely
on the PeerConnection getStats API to determine when data packets have both been sent
and received~\footnote{packetsSent and packetsReceived in the ssrc stats type}.
The results for the local and remote gateway settings using the fast laptop and wired network
setup can be seen in Figure~\ref{fig:connectiontime}.

We can see that stream setup time is not effected by our gateway, but media setup incurs 2-5
seconds of overhead compared to a direct connection for local and remote gateways respectively.

\subsection{Scalability}
We have designed the gateway to be shared among clients on the same local network.
Two requirements result from that.
First, the gateway should be easy to reuse between clients, and
second it should handle concurrent streams without too much overhead.
Since the second requirement relies on the first we only focus on showing the
evaluating the second here. Based on our design the first requirement is also
trivially met, as we create separate leased tunnels (essentially ports) for each session.

\begin{figure}
\hfill
\subfigure[RTT]{\includegraphics[width=4.1cm]{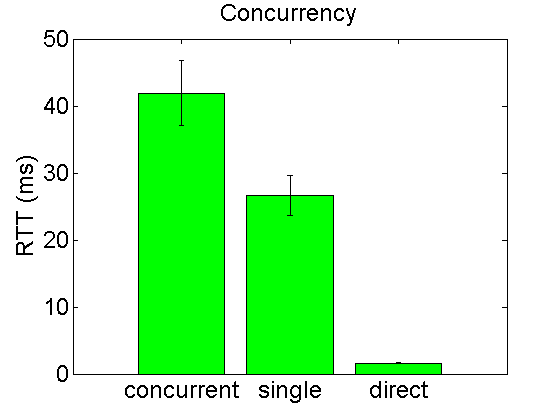}}
\hfill
\subfigure[Jitter]{\includegraphics[width=4.1cm]{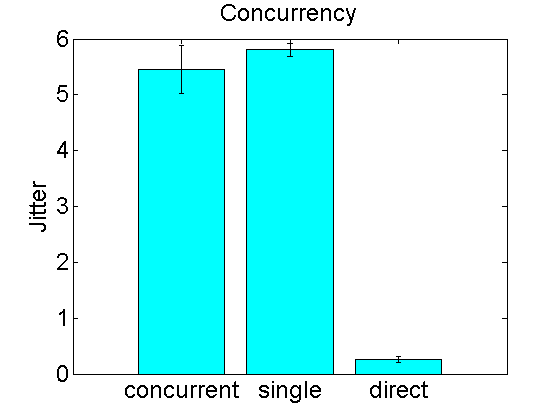}}
\hfill
\caption{Latency in the concurrent and single stream setups compared to direct link sessions involving
three laptops and a desktop on the same wired network.}\label{fig:concurrency}
\end{figure}

To evaluate the overhead of running multiple concurrent streams in the same LGAT and CGAT
services we set up an experiment involving four machines, the desktop, and the fast and slow
laptop from the previous experiment and another new laptop. In the direct connection baseline
experiment we let the fast and slow laptops communicate peer-to-peer at the same time as the 
desktop and the new laptop communicate peer-to-peer. We chose this setup since the new laptop
is slower than both the slow and the fast laptops and the desktop is the fastest thus keeping
the average performance of the two endpoints roughly the same in the two sessions.
Now to evaluate concurrent stream overhead in a single gateway we let the new laptop use
the fast laptop's gateway (LGAT), and the slow laptop use the desktop's gateway (LGAT).
The desktop and the fast laptop both use their own local gateways (LGAT). We note that this leads
to both streams sharing both LGATs and also the same single CGAT. We now start both sessions, and
let them run concurrently for about 7 minutes to extract our benchmark latency values from the desktop
peer and its session with the new laptop. Note, that we only use the session between the fast and slow laptop
as overhead and do not measure it directly. After the 7 minutes have passed we stop the session
between the old and new laptops and observe how the performance instantly picks up for the other
session that we keep alive.
The results keen be seen in Figure~\ref{fig:concurrency}.

We note that there is a 20ms drop in RTT when the concurrent stream stops.
Given that the absolute performance of a single stream is also about 20ms
we can extrapolate the potential number of streams our gateway can handle
given a RTT requirement. As a rule of thumb at about 200ms RTT the session
starts feeling like it is lagging. This means that our gateway could handle
up to 10 concurrent sessions assuming no other contentions occur. Given that
it is easy to load balance both between LGAT and CGAT services we believe this
to be a reasonable overhead. Note also that an LGAT/CGAT can
be shared among a much high number of users, since all users would not use the
gateway at the same time.

\section{Conclusions}\label{sec:conclusion}
We have presented a solution for tunneling WebRTC data plane traffic efficiently
with minimal setup through restricted networks having strict firewall policies.

Our solution relies on intercepting key parts of the WebRTC JSEP session
establishment protocol and using local network gateways that can multiplex
traffic from multiple concurrent streams efficiently without ``leaking'' any
WebRTC traffic across the firewall except through a trusted port such as port 80.

The advantage of our solution compared to existing work is that the traffic is relayed
and all the dynamic port allocation is done on the local network, while still allowing the
traffic to be re-routed to arbitrary networks. It would even be trivial to load balance our 
traffic on the fly by simply relaying the packets through a different central gateway. 
None of the communicating browsers would have to know. Given the use of SRTP 
encryption of the payloads there is no risk of eavesdropping middlemen.

We have shown that the latency impact of our gateway is acceptable to the user
experience up to about 10 concurrent streams going through the same gateway 
(given a per-stream RTT overhead of about 20ms).
The WebRTC runtime in the browser is quite CPU intensive in general, so the browser that runs the
end-user application needs to be powerful enough to be able to avoid additional latency. 
Our gateway can be run on a local device collocated with the browser to minimize network traffic,
but the CPU impact of this setup is significant. So it needs to be a powerful machine to avoid
latency. This performance impact and the additional configuration effort is the reason 
why we recommend
deploying dedicated machines on the local network to run our gateways. They may easily 
be replicated and load-balanced. Providing automated gateway selection, local and
central, could improve the scalability significantly, but was out of scope
for the work presented here.

Our experiments also showed that the additional session establishment overhead
is acceptable, in the order of 1-5 seconds. Here, there is a small advantage
to having a local host gateway. 

Future work includes testing our gateway with more browsers as the WebRTC
PeerConnection API is being supported by more vendors, and investigating
the feasibility of hiding our implementation behind the TURN protocol
to avoid custom Javascript code to redirect the traffic during
the JSEP handshake. Embedding an additional Javascript library does not
have much impact on the application, though, since applications would need to
embed a custom WebRTC library in any case as the signalling plane 
implementation is application specific. 

We are also continuing work on deploying this solution in our
home-care project at hospitals and with patients.

\section*{Acknowledgement}\label{sec:acknowledgement}
This work was supported by the Swedish research fund VINNOVA in the program {\it Challenge Driven Innovation} under contract 2011-02796.


\bibliographystyle{abbrv}
\bibliography{palcomweb}  
%
\end{document}